\documentclass[10pt,journal,compsoc]{IEEEtran}
\usepackage{graphicx}
\usepackage[ruled]{algorithm2e}
\usepackage[utf8]{inputenc}
\usepackage{subfigure}
\usepackage{multirow}
\usepackage{tabulary}
\usepackage{amsthm}
\usepackage{amssymb}
\usepackage{algorithmic}
\usepackage{enumitem}
\usepackage{epstopdf}
\usepackage[justification=centering]{caption}
\usepackage{amsmath}

\ifCLASSOPTIONcompsoc
  \usepackage[nocompress]{cite}
\else
  \usepackage{cite}
\fi
\ifCLASSINFOpdf
\else
\fi
\begin{document}
%
\title{Comprehensive Privacy Analysis on Federated Recommender System against Attribute Inference Attacks}
%
%
%
%

\author{Shijie Zhang, Wei Yuan and Hongzhi Yin, \textit{Senior Member, IEEE} 

\IEEEcompsocitemizethanks{\IEEEcompsocthanksitem S. Zhang, W. Yuan and H. Yin are with the school of Information Technology and Electrical Engineering, The University of Queensland, Brisbane, Australia,
E-mail: shijie.zhang@uq.edu.au, w.yuan@uq.edu.au and h.yin1@uq.edu.au}
\thanks{Hongzhi Yin is the corresponding author.}
}


%
%

\markboth{Journal of \LaTeX\ Class Files,~Vol.~14, No.~8, August~2015}%
{Shell \MakeLowercase{\textit{et al.}}: Bare Advanced Demo of IEEEtran.cls for IEEE Computer Society Journals}
%



\IEEEtitleabstractindextext{%
\begin{abstract}

In recent years, recommender systems are crucially important for the delivery of personalized services that satisfy users' preferences. With personalized recommendation services, users can enjoy a variety of recommendations such as movies, books, ads, restaurants, and more. Despite the great benefits, personalized recommendations typically require the collection of personal data for user modelling and analysis, which can make users susceptible to attribute inference attacks. Specifically, the vulnerability of existing centralized recommenders under attribute inference attacks leaves malicious attackers a backdoor to infer users' private attributes, as the systems remember information of their training data (i.e., interaction data and side information). An emerging practice is to implement recommender systems in the federated setting, which enables all user devices to collaboratively learn a shared global recommender while keeping all the training data on device. However, the privacy issues in federated recommender systems have been rarely explored. In this paper, we first design a novel attribute inference attacker to perform a comprehensive privacy analysis of the GCN-based federated recommender models. The experimental results show that the vulnerability of each model component against attribute inference attack is varied, highlighting the need for new defense approaches. Therefore, we propose a novel adaptive privacy-preserving approach to protect users' sensitive data in the presence of attribute inference attacks and meanwhile maximize the recommendation accuracy. Extensive experimental results on two real-world datasets validate the superior performance of our model on both recommendation effectiveness and resistance to inference attacks.   
\end{abstract}

}

\maketitle

\IEEEdisplaynontitleabstractindextext

%
\IEEEpeerreviewmaketitle

\ifCLASSOPTIONcompsoc
\IEEEraisesectionheading{\section{Introduction}\label{sec:introduction}}
\else
\section{Introduction}
\label{sec:introduction}
\fi
In online services, the demand for recommender systems has increased more than ever before, due to their success in alleviating the problem of information overload by filtering vital information out of a large volume of data to efficiently deliver personalized contents and services for users~\cite{chen2019air, yin2019social, chen2020try}. It is no wonder, these recommenders, if set up and configured properly, can significantly contribute to revenues as well as user experience. In recent years, various recommendation algorithms have been proposed and achieved immense success in practical applications. Collaborative filtering (CF) based recommender systems, which make recommendations by utilizing users' historical interaction data, are widely deployed in the online platforms, for the fact that they are effective and efficient. More recently, deep learning-based recommender systems have demonstrated advantageous effectiveness by advancing the representation learning capability and producing high-quality recommendations~\cite{ying2018graph,yu2020enhance, yin2015joint, yin2016spatio}.



\begin{figure}
    \centering
    \includegraphics[scale=0.38]{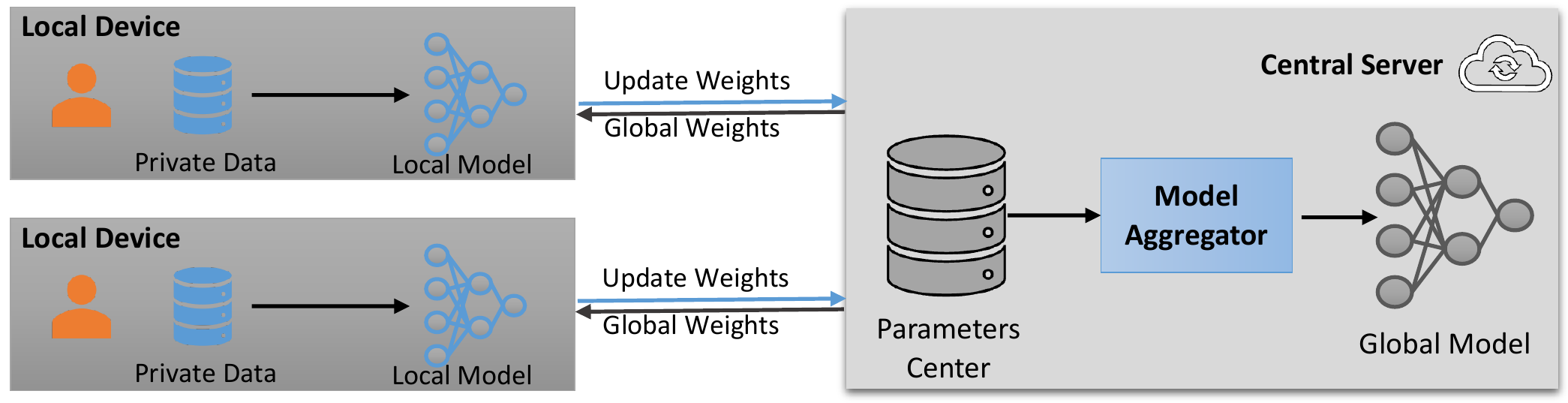}
    \caption{Architecture of a typical Federated Recommender System.}
    \vspace{-2em}
    \label{fig:fedrec}
\end{figure}

Despite the promising effectiveness, tremendous privacy concerns on recommender systems are raised in recent years. On one hand, to boost the recommendation performance, especially for fresh (i.e., cold-start) customers, these systems hungrily collect various side information data (a.k.a. user attributes or contexts) to better infer users' preferences~\cite{adomavicius2011context,zhang2020gcn}. When registering accounts, users are required to complete questionnaires about their personal demographics (i.e., age and gender) to facilitate user profiling. Once intercepted by malicious third parties, the leakage of users' sensitive information can be catastrophic. On the other hand, recent research indicates that even users' unpublished private information can be inferred via their interaction history with high confidence~\cite{weinsberg2012blurme,kosinski2013private,calandrino2011you}. Such personal information includes but not limited to age, gender, political orientation, health, financial status etc. Much worse, the attackers can utilize the inferred attributes to link user accounts across multiple platforms and break anonymity~\cite{shu2017user,goga2013exploiting}. Such attack is termed as attribute inference attack~\cite{gong2016you}, where the malicious attackers can be data brokers, cyber criminals, advertisers, etc. An example is that~\cite{naranyanan2008robust} successfully deanonymizes Netflix users by utilization of the publicly accessible IMDb user profiles. Accordingly, the personalized recommendation results can also cause privacy leakage since they imply users' interests and even their sensitive attributes~\cite{10.1145/3442381.3449813, beigi2020privacy}. In previous work~\cite{10.1145/3442381.3449813}, a novel differentially private graph convolutional network named GERAI is proposed to address aforementioned privacy issues. Specifically, the graph convolutional networks (GCNs)~\cite{hamilton2017inductive,kipf2016semi} is adopted as the main building block, since it is capable of jointly exploiting the user-item interactions and the rich side information of users. Then, to provide a privacy guarantee, the authors design a novel dual-stage perturbation paradigm with differential privacy, which makes the recommendations less dependent on users' sensitive data, avoiding privacy leakage in recommendation results. Unfortunately, despite its success, this centralized recommendation paradigm still inevitably leads to increasing risks to user privacy, since user data stored on the central server might be accidentally leaked or misused. 

In light of privacy issues in centralized recommender systems, there has been a recent surge in decentralized recommender systems, where federated learning (FL) \cite{wu2021fedgnn,muhammad2020fedfast, wang2021fast} becomes one of the most representative frameworks in the development of privacy-preserving systems. Specifically, a federated recommender shown in Figure~\ref{fig:fedrec} allows users' data safely hosted on their personal devices, and the shared global recommender is collaboratively trained in a multi-round fashion by collecting a batch of locally updated models to the central server for parameters update. To avoid privacy concerns, the server is designed to have no access of each client's local data and training process. As a result, such federated recommenders seem to be regarded as ‘safe’ towards attribute inference attacks. Though federated learning framework can achieve comparable recommendation results without sharing users' sensitive data, recent works show that it is yet to provide a privacy guarantee of training data~\cite{papernot2016semi}. Specifically, model parameters uploaded during the training process provide a chance for inference attacks, since the well-trained parameters can remember local data information.~\cite{nasr2019comprehensive,truex2019demystifying,chai2020secure} have studied that deep learning models in the federated setting are susceptible to membership inference attacks, where the attacker is able to infer the samples used to train the model.

\begin{figure}[t!]
\centering
\begin{tabular}{cc}
\multicolumn{2}{c}{ \includegraphics[scale=0.42]{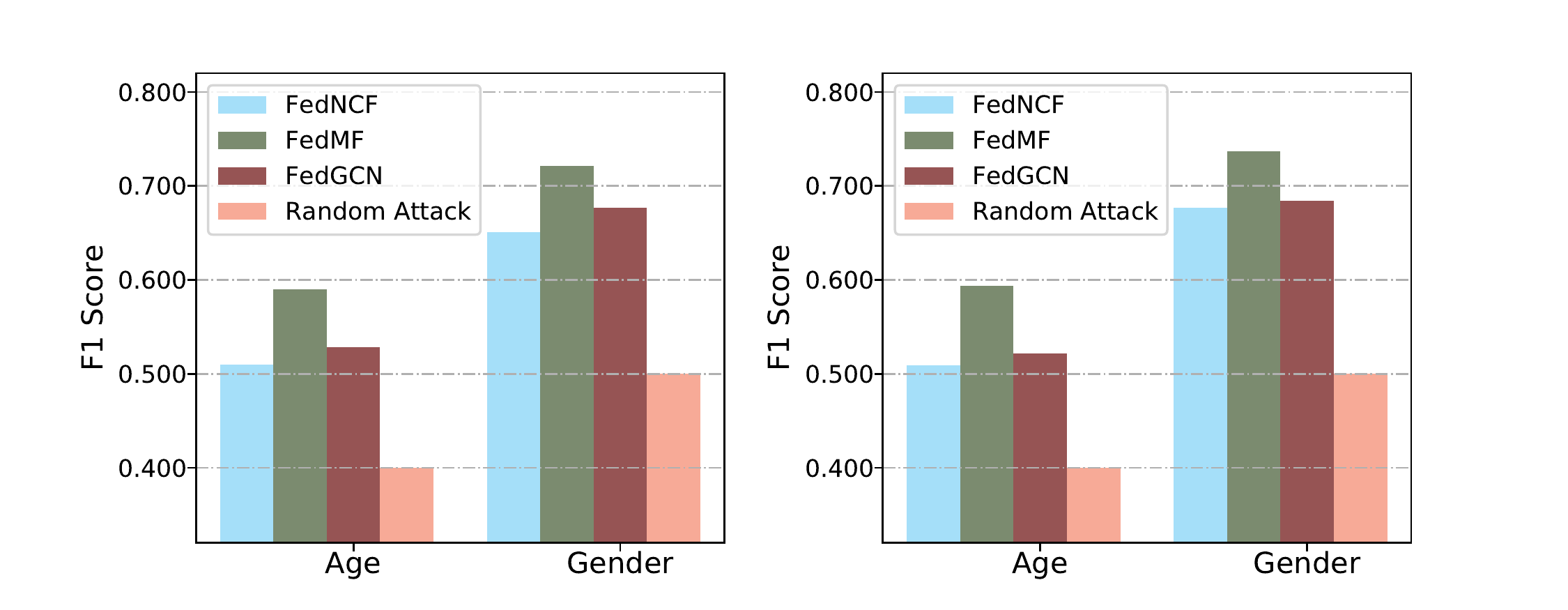}}\\
\hspace{4.5em}(a)ML-100K&\hspace{3.5em}(b)ML-1M\\

\end{tabular}
\vspace{-0.5em}
\caption{Attribute inference attack results of popular federated recommenders on ML-100K and ML-1M. }
\label{fig:attack}
\vspace{-1em}
\end{figure}

Although membership inference attacks have been studied in the federated setting~\cite{nasr2019comprehensive,truex2019demystifying,chai2020secure}, the vulnerability of federated recommender against attribute inference attacks (i.e., users' attributes) remains unexplored. To validate whether federated recommender paradigm is susceptible to attribute inference attacks, we make the first attempt to infer users' private attributes through the uploaded parameters in the federated setting. Figure~\ref{fig:attack} demonstrates the F1 Score achieved by the attribute inference attacker (refer to details in Section~\ref{sec:attacker}) on three well-trained federated recommenders (MF~\cite{ammad2019federated}, NCF~\cite{zhang2022pipattack} and GCN~\cite{kipf2016semi}). To show the significance of attribute inference from local model parameters, we build a random guessing classifier named Random Attack as a null model. From the results shown in Figure~\ref{fig:attack}, we observe significant differences between the random attack and the inference attack based on local model parameters. It can be concluded that the shared model parameters in the federated learning process significantly reveal users' attribute information, demonstrating the demand for advanced defenses against attribute inference attacks to federated recommender systems. Recently, local differential privacy (LDP) has become a gold standard for 
providing protection guarantee of local model parameters in the federated setting.~\cite{qi2020privacy, wu2021fedgnn, nguyen2016collecting, wang2019collecting} successfully apply LDP mechanism in federated deep learning tasks to transform the local model parameters into a noisy version at each user device before being uploaded to the central server. Despite its success in many applications against membership attacks, it has been rarely studied to protect users' attribute information in the federated recommendation.


These aforementioned limitations motivate us to propose a novel privacy-aware federated recommender system that significantly improve both recommendation effectiveness and robustness against attribute inference attacks. However, how to apply LDP technique in the federated recommender systems faces tremendous challenges. Setting an appropriate value for the privacy budget is crucial for the utility of the attack-resistance recommender system in the federated setting. A low budget value (i.e., noise factor) can result in a high success rate of attribute inference attack since the noise added into model parameters are negligible and ineffective, while a high value will inevitably destroy model utility. For simplicity, existing works~\cite{qi2020privacy, wu2021fedgnn} just fix a constant DP budget for all model parameters. However, designed for modelling nonlinear relations between users and items, federated deep recommender models have multiple components and layers, and their model parameters exhibit large variance, thus the vulnerability of each component or even each layer to the attribute inference attacks is different. Therefore, treating all components/layers under the same privacy protection mechanism results in unavoidable excessive utility loss.    

In this paper, we perform a comprehensive privacy analysis of each component of the GCN-based federated recommender model, and the results show varied vulnerabilities of these components against attribute inference attacks. Specifically, we divide model parameters into three main parts based on functionality, namely User Component, Item Component and MLP. To achieve optimal privacy strength without sacrificing much recommendation accuracy, we design a novel adaptive LDP mechanism named APM that can automatically adjust the utility loss of each component when defending attribute inference attacks. To conclude, we highlight our main contributions as follows:
\begin{itemize}
    \item To the best of our knowledge, we are the first to present a comprehensive privacy analysis of federated recommender systems under attribute inference attacks. Our study reveals that well-trained recommenders are significantly susceptible to attribute inference attacks even in a decentralized environment.
    \item In order to address growing privacy concerns in the federated recommendation context, we design a novel adaptive privacy-aware mechanism to guard users’ sensitive data against attribute inference attacks without sacrificing high-quality recommendation results. Our model innovatively takes advantage of the inherent informative bias to reduce overall utility loss in defending attribute inference attacks. 
    \item Extensive experiments are conducted on two real-world datasets, and the results demonstrate the superior performance of our solution. Furthermore, compared with all baselines, the results show that our model provides a strong privacy guarantee with less compromise on the recommendation accuracy. 
\end{itemize}

\section{PRELIMINARIES}
In this section, we first revisit the key definitions that are frequently used in this paper and then formulate our problem. Note that we use bold lowercase (e.g., $\textbf{a}$) to denote vectors, and use bold uppercase (e.g., $\mathbf{A}$) to denote matrices. All sets are written in squiggle uppercase letters (e.g., $\mathcal{A}$).

\textbf{Differential Privacy.} Differential privacy (DP) can provide a rigorously mathematical guarantee in the machine learning tasks. The notion of differential privacy was first introduced by~\cite{dwork2014algorithmic} and it can be utilized to defend against malicious attackers that infer useful information from a target model (e.g., outputs and structure of the model). Given a privacy parameter $\epsilon > 0$, the $\epsilon-$differential privacy ($\epsilon-$DP) is defined as follows:

\textit{Definition 2.1}. ($\epsilon-$Differential Privacy) A randomized mechanism $f: \mathcal{D} \rightarrow \mathcal{R}$ with domain $\mathcal{D}$ and range $\mathcal{R}$, and it satisfies $\epsilon-$DP if: 
\begin{equation}\label{eq:DP}
    Pr[f(\mathcal{D}) \in O] \leq exp(\epsilon)Pr[f(\mathcal{D}')\in O],
\end{equation}
where $Pr[\cdot]$ represents probability, $\mathcal{D}$ and $\mathcal{D}^{'}$ are two adjacent datasets differing on only one data instance, and $O\subseteq \mathcal{R}$ denotes any subsets of possible output values. Eq.(\ref{eq:DP}) implies that the probability of output distribution $f(\mathcal{D})$ is at most $exp(\epsilon)$ times smaller than that of $f(\mathcal{D}')$. On this basis, $f(\cdot)$ is not overly dependent on any individual data record, providing each instance roughly the same privacy. In the federated setting, the central server updates global model by aggregating the collected model parameters from a group of local devices. To defend against malicious attacks that infer user's private attributes via its device's updates, each local device should first perturb model parameters $\Theta$ by directly adding noise, and then the noised version of $\Theta^*$ is updated to the central server instead of original parameters. Hence, the model parameter generated by each local device is treated as a singleton dataset, and we require the random perturbation mechanism $f(\cdot)$ to perform by the local devices, not by the central server. Specifically, we introduce $\epsilon-$local differential privacy ($\epsilon-$LDP) which is a special case of differential privacy: 

\textit{Definition 2.1}. ($\epsilon-$Local Differential Privacy) A randomized mechanism $f(\cdot)$ satisfies $\epsilon-$LDP if and only if for any two input data $\Theta$ and $\Theta'$, we have:
\begin{equation}\label{eq:LDP}
    Pr[f(\Theta) = \Theta^{*}] \leq exp(\epsilon) \cdot Pr[f(\Theta') = \Theta^{*}]
\end{equation}
where $\Theta^{*}$ denotes the output of $f(\cdot)$. The lower $\epsilon$ provides stronger privacy but may result in severe performance drop of a federated model as each local model is heavily perturbed. Hence, $\epsilon$ determines privacy budget that controls the trade-off between privacy and model utility. With the privacy guarantee from LDP, an external attacker who collects $\Theta^{*}$ (e.g., perturbed model parameters) cannot accurately estimate the true data is $\Theta$ or $\Theta'$, and thus the sensitive information is obfuscated. 


\textbf{Federated Recommender Systems.}
Let $\mathcal{V}$ and $\mathcal{U}$ denote the sets of $N$ items and $M$ users, respectively. Each device used by an individual user $u\in \mathcal{U}$ has a local training dataset $\mathcal{D}_{u}$ that consists of implicit feedback tuples $(u, v, r_{uv})$, where $r_{uv} = 1$ if $u$ has interacted with item $v \in \mathcal{V}$ (i.e., a positive instance), otherwise, we set $r_{uv}$ to $0$ (i.e., a negative instance). Due to the large number of unobserved interactions, we use a sample ratio of $1:q$ to downsample the negative instances for each user $u$. We use $\mathcal{N}(u)$ to denote the set of items visited by $u$. Additionally, each user $u$ privately preserves a dense input vector $\mathbf{x}_u\in \mathbb{R}^{d^1}$ in which each element represents either $u$'s private attribute $s\in \mathcal{S}$ or a extracted statistical feature $s\in \mathcal{S}'$ based on $\mathcal{D}_{u}$. Not that each categorical feature (i.e., age and gender in our case) is represented by one-hot encoding in $\mathbf{x}_u$. The federated recommender system aims to train a global recommender across multiple decentralized user devices that hold local private data (i.e., $\mathbf{x}_u$ and $\mathcal{D}_u$), without direct access to them:  
\begin{equation}
   \underset{\Theta_u}{\mathrm{argmin}}\sum_{u \in \mathcal{U}} \mathcal{L}^{rec}(\mathcal{D}_u, \Theta_u)
\end{equation}
 where $\mathcal{L}^{rec}(\cdot)$ is a loss function and $\Theta_u$ represents all the trainable parameters of $u$'s local recommender. For notation convenience, we use $\Theta$ to represent the recommender system.

\textbf{Task 1.} For each user $u \in \mathcal{U}$, given its local dataset $\mathcal{D}_u$ and user feature vector $\textbf{x}_u$, we aim to learn a privacy-preserving federated recommender system, in which malicious attackers are unable to infer user's private attributes (i.e., gender and age in our case) via $u$'s uploaded model parameters with high confidence.
\section{Attribute INFERENCE ATTACKS}
\begin{figure*}
    \centering
    \includegraphics[scale=0.55]{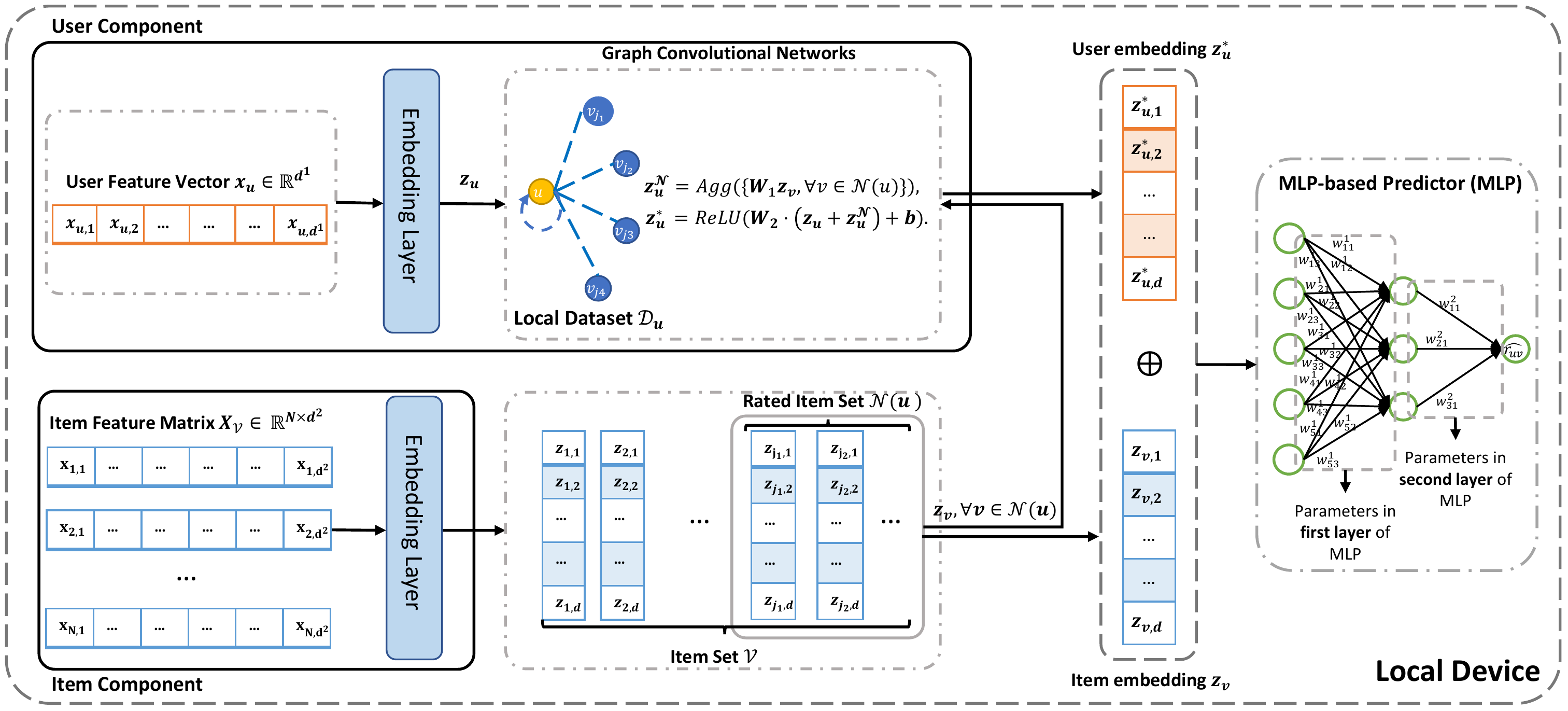}
    \caption{Overview of Local model in each device.}
    \label{fig:localrec}
    \vspace{-1.5em}
\end{figure*}
\subsection{Base Recommender}\label{sec:base_rec}
Federated learning appears to be compatible with various
latent factor models. The advantage of FL that users' data host on their local devices makes it attractive for developing privacy-preserving models. In this paper, we extend the DSSM~\cite{huang2013learning} that is a widely used backbone for centralized recommendation to a federated one named FedRec.  The local FedRec framework is shown in Figure~\ref{fig:localrec}, which consists of three key components to generate recommendations, namely user component, item component and MLP. Notably, almost all federated recommendation systems designed for top-k recommendations follow this architecture, such as\cite{zhang2022pipattack, wang2021fast, wu2021fedgnn, yuan2023federated, wu2022fedattack,lin2020meta,chai2020secure, qi2020privacy, muhammad2020fedfast}, since it is generic and can be easily extended to most advanced recommenders  by simply adopting different feature modeling layers in user or item components. We employ a GCN~\cite{ hamilton2017inductive} layer as the key building block to learn user embeddings in FedRec, since it is advantageous in capturing local structure information of the user-item interaction data and user's side information in a unified way. To guarantee user privacy, the neighbor set $\mathcal{N}(v)$ of item $v$ is highly sensitive and restricted, and thus GCN layer cannot be applied to item embeddings. In what follows, we will introduce the design details of each component in FedRec.

In user $u$'s local device, given local dataset $\mathcal{D}_u$ and we assume that a feature vector $\mathbf{x}_u$ associated with $u$ is available. Note that we use $\mathbf{z}_u$ and $\mathbf{z}_v$ to denote the latent representations of users and items in the same latent space, respectively. Specifically, $\textbf{z}_u, \textbf{z}_v$ can be initialized as follows:
\begin{equation}
	\textbf{z}_u = \textbf{E}_{\mathcal{U}}\textbf{x}_u, \,\,\textbf{z}_v = \textbf{E}_{\mathcal{V}}\textbf{x}_v,
\end{equation}
where $\textbf{x}_u \in \mathbb{R}^{d^1}$ is user $u$'s raw feature vector and $\textbf{E}_{\mathcal{U}} \in \mathbb{R}^{d\times d^1}$ is the user feature transformation matrix. $\textbf{x}_v \in \mathbb{R}^{d^2}$ and $\textbf{E}_{\mathcal{V}} \in \mathbb{R}^{d\times d^2}$ represent item feature vector and item feature transformation matrix. To ensure our model's generalizability, each item feature vector $\textbf{x}_v$ is initialized using randomized values as we do not assume the availability of item features.

In each forward iteration, to learn the embedding of user $u$, GCN firstly computes its embedding by iteratively aggregating information from its first-order neighbors, i.e., items interacted with $u$: 
\begin{equation}
	\mathbf{z}^{\mathcal{N}}_u= Agg(\{\mathbf{W}_1 \mathbf{z}_v, \forall v\in \mathcal{N}(u)\}),
\end{equation}
where $\mathbf{W}_1$ is trainable weight matrix and $Agg(\cdot)$ is the aggregation function which aggregates the neighborhood information $\mathbf{z}_{v}$ into a unified vector representation. In the experiment, we adopt the average aggregation function for simplicity.

Then, the user's current representation $\mathbf{z}_u$ is added to the aggregated neighborhood vector $\mathbf{z}_u^{\mathcal{N}}$, and then being fed through a fully connected layer to form an updated user embedding $\textbf{z}_u^{*}$: 
\begin{equation}\label{eq:convolution}
    \mathbf{z}_{u}^{*} = ReLU(\mathbf{W}_2 \cdot (\mathbf{z}_u + \mathbf{z}_u^{\mathcal{N}}) + \mathbf{b}),
\end{equation}
where $ReLU(\cdot)$ denotes the rectified linear unit for nonlinearity, and $\mathbf{W}_2$ and $\mathbf{b}$ are weight matrix and bias vector. 


On the item side, as the neighbor set of each item (i.e., the set of users who interact with the item) is unavailable in the federated setting, we do not perform convolution operation when learning item representations. Therefore, the parameters in user component and item component can be represented as $\Theta^{\mathcal{U}} = \{\textbf{E}_{\mathcal{U}}, \mathbf{W}_1, \mathbf{W}_2, \mathbf{b}\}$ and $\Theta^{\mathcal{V}} = \{\textbf{E}_{\mathcal{V}}\}$.

Afterwards, in user $u$'s local device, the ranking score $\hat{r}_{uv}$ for an arbitrary item $v\in \mathcal{N}(u)$ can be predicted by $u$'s local recommender. To achieve this, local recommender first concatenate $u$'s and target item $v$'s current embeddings, and then adopt a $L$-layer perceptron network (MLP) $MLP(\cdot)$ to model the user-level interactions and estimate $\hat{r}_{uv}$:

\begin{equation}
\label{eq:rec}
    \hat{r}_{uv} = \sigma(MLP(\mathbf{z}_u^* \oplus \mathbf{z}_v)),
\end{equation}
where $\textbf{z}_u^*, \textbf{z}_v \in \mathbb{R}^d$ are user and item embeddings respectively, $\oplus$ is the concatenation operation, and $\sigma$ is the sigmoid function that rectifies the ranking score to the range $[0, 1]$.
\vspace{-0.5em}
\subsection{Federated Learning Protocol.}\label{sec:fl} 
As shown in Algorithm~\ref{alg:FedRec}, FedRec aims to train a shared recommender with a central server by coordinating individual user devices to train local models based on private dataset $\mathbf{\mathcal{D}}_{u}$ and user features $\mathbf{x}_u$. To train FedRec, each user's model is optimized locally with a distance-based loss function:
\begin{equation}
\resizebox{.9\linewidth}{!}{$
    \mathcal{L}^{rec} = - \sum_{(u,v, r_{uv})\in \mathcal{D}_u} r_{uv}\log \hat{r}_{uv} + (1-r_{uv})\log(1-\hat{r}_{uv}),
    $}
\label{eq:rating}
\end{equation}
where $\hat{r}_{uv} \in [0,1]$ is obtained via Eq (\ref{eq:rec}). Notably, we adopt cross-entropy loss to minimize the difference between $\hat{r}_{uv}$ and the ground truth ${r}_{uv}$. With the user-specific loss $\mathcal{L}^{rec}$ computed, we can obtain updated parameters $\Theta^t_u$ of $u$'s local model. Specifically, at each epoch $t$, a subset of users $\mathcal{U}^t$ are randomly drawn, and each selected local device should download the latest global recommender $\Theta_t$ and then update its local recommender on $\mathcal{D}_u$. Then, each device uploads its updated local model parameters to the central server. Once the central server receives all local model parameters submitted by $|\mathcal{U}_t|$ users, it aggregates the collected parameters to facilitate global recommender update. Specifically, FedRec follows the commonly used FedAvg protocol \cite{mcmahan2017communication} to update global recommender $\Theta_{t+1}$:
\begin{equation}
\Theta^{t+1} = \sum_{u \in \mathcal{U}_t} \Theta_u^{t}.
\end{equation}
The training continues iteratively until the global recommender achieves convergence criteria. Unlike centralized recommenders, FedRec collects only each user's local model parameters instead of personal data (i.e., $\mathcal{D}_u$ and demographics $\textbf{x}_u$), and local model parameters are not directly shared across users. 
\vspace{-1em}
\subsection{Attribute Inference Attack}
\label{sec:attacker}

\begin{algorithm}[t]
\caption{Procedures for Training FedRec}
\KwIn{The number of global rounds $\mathcal{T}_g$ and local Epochs $\mathcal{T}_l$, sampled clients $\mathcal{U}^t$ at each time $t$, initialized model parameters $\Theta$ and local learning rates $\mu_1$;}\
\For{$t\leq \mathcal{T}_g$}{
\textbf{Local Training Process for $u \in \mathcal{U}^t$:}\\
Initialization: $\Theta^t_u \leftarrow \Theta$ \;
\For{$t_2\leq \mathcal{T}_l$}
{
Draw a minibatch $\mathcal{B}$ \;
\For{$\mathcal{B}_i \in \mathcal{B}$}{
$\Theta^t_u \leftarrow \Theta^t_u - \mu_1 \frac{\partial \mathcal{L}^{rec}(\mathcal{B}_i, \Theta^t_u)}{\partial \Theta_u^t} $\;
}
$t_2 \leftarrow t_2 +1$\;
}
Send $\Theta^t_u$ to Server\;
\textbf{Model Aggregating process:}\\
Update the global parameters $\Theta$ as:\\
$\Theta^{t+1} = \frac{1}{|\mathcal{U}^t|} \sum_{u \in \mathcal{U}^t} \Theta_u^{t}$\;
$\Theta \leftarrow \Theta^{t+1}$\;
$t \leftarrow t + 1$\;
}
\label{alg:FedRec}
\end{algorithm}
\setlength{\textfloatsep}{1em}

\begin{figure*}[t!]
\centering
\begin{tabular}{cccc}
\multicolumn{4}{c}{ \includegraphics[scale=0.46]{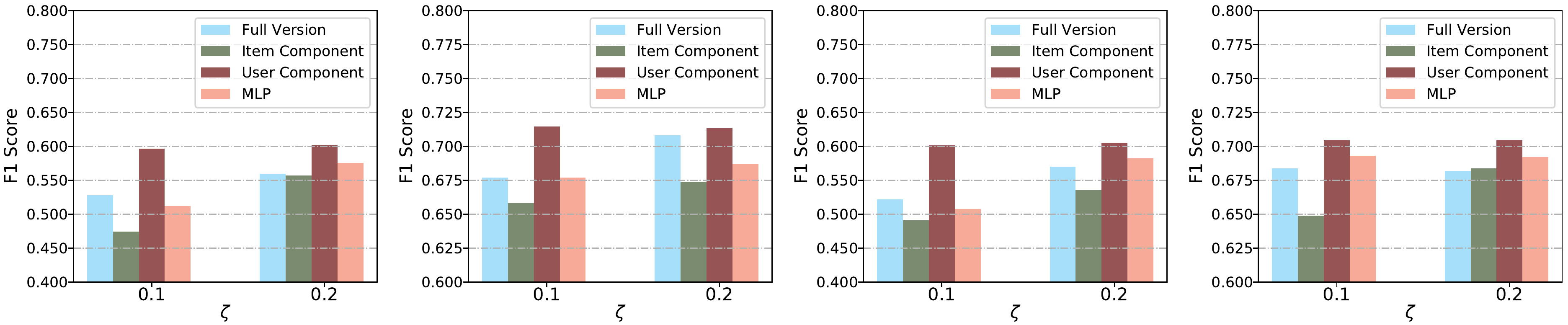}}\\
\hspace{3.2em}(a)Age on ML-100K&\hspace{2.5em}(b)Gender on ML-100K&\hspace{3em}(c)Age on ML-1M&\hspace{2.8em}(d)Gender on ML-1M\\
\multicolumn{4}{c}{ \includegraphics[scale=0.46]{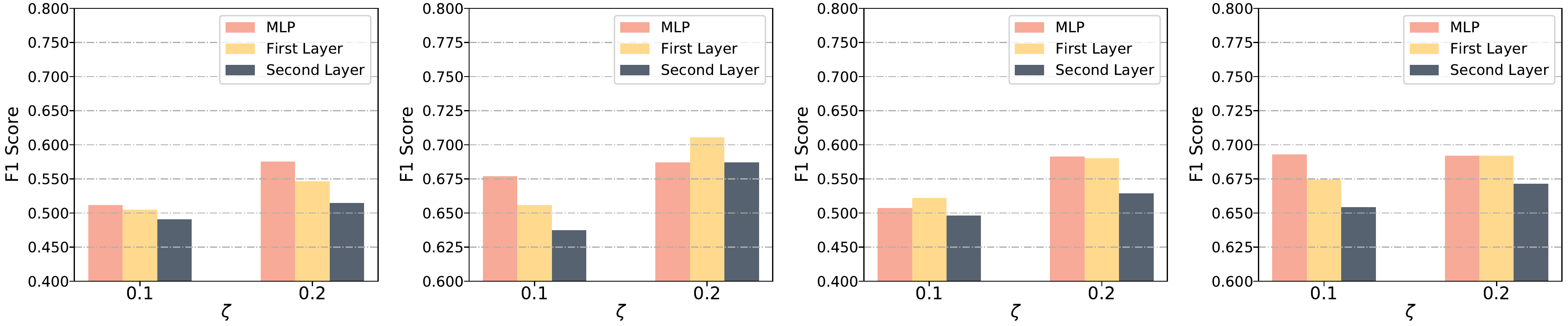}}\\
\hspace{3.2em}(a)Age on ML-100K&\hspace{2.5em}(b)Gender on ML-100K&\hspace{3em}(c)Age on ML-1M&\hspace{2.8em}(d)Gender on ML-1M\\
\end{tabular}
\vspace{-0.5em}
\caption{Vulnerability of each component in FedRec defending against attribute inference attack w.r.t. $\zeta$. }
\label{fig:att_f1}
\vspace{-1em}
\end{figure*}
Generally, federated recommender system assumes that the central server is trusted and works under non-adversarial setting. In real-life scenarios, the server is honest-but-curious, which means server is curious in inferring the information of individuals via received updates but honest in processing the updates, leaving a backdoor to estimate users' private attributes. To perform attribute inference attacks on FedeRec, traditional attack approaches designed for centralized recommenders have limited success to attack federated models. The main reason is that the prior knowledge (e.g., all user-item interactions and recommendation results) that is requisite for the malicious attackers~\cite{10.1145/3442381.3449813, beigi2020privacy}, cannot be obtained in the federated setting as such information is kept privately at the user side. Furthermore, the federated setting substantially restricts the knowledge acquisition of an attribute inference attack model, and we summarized the accessible prior knowledge as follows:
\begin{itemize}[leftmargin=*]
	\item[I.] The adversary can access user $u$'s local model parameters $\Theta_u^t$ at any iteration $t$.
	\item[II.] The adversary knows sensitive attributes of a small group of users $\mathcal{U}_{adv}$ who may cooperate with the untrusted server due to financial incentives, or general registers who are willingness to share their information with the online platforms. 
\end{itemize}
With the updated model weights $\Theta^t$, we can obtain $\Delta \Theta = \frac{\Theta^{t-1} - \Theta^{t}}{\mu_1}$, which can be regarded as the gradient of model parameters over one epoch of SGD optimization method. $\Delta \Theta^t$ reflects how much each parameter has to change and contains sufficient information of the training dataset. As a result, the adversary has a dataset $\mathcal{D}_{adv} = \{(\Delta \Theta_u^t, y_u)| \forall u \in \mathcal{U}_{adv} \}$. To construct an attribute inference attack model, the adversary needs to find the meaningful mapping between the model's updated parameters and user attributes. Given the dataset $\mathcal{D}_{adv}$, the most straightforward way of learning such relationship is to train the attack model in a supervised way, and use it to attack the rest users whose attributes are unshared. Then, suppose there are $C$ classes of target attribute, the attribute inference attacker (AIA) $f_{adv}(\cdot)$ is a three-layer deep neural network that inputs a model update gradient $\Delta \Theta_u$, and outputs a $C$-dimensional vector $\hat{\mathbf{y}}$ in which each element $\hat{\mathbf{y}}[c] \in \hat{\mathbf{y}}$ $(c = 1, 2,..., C)$ denotes the probability that user $u$ is classified to label $c$. We train $f_{adv}(\cdot)$ with cross-entropy loss on all training dataset $\mathcal{D}_{adv}$ :
\begin{equation}
     \mathcal{L}_{adv} = - \sum_{\forall{u \in \mathcal{U}_{adv}}}\sum_{c=1}^{C} \textbf{y}[c]\log \hat{\textbf{y}}[c],
\end{equation}
where $\mathbf{y} = \{0, 1\}^C$ is the one-hot label and the hidden dimension is set to $100$ and $30$. To evaluate FedRec's robustness against the proposed attack model, we randomly choose $\zeta$ of users in $\mathcal{U}^{t}$ as $\mathcal{U}_{adv}$ to train our attacker, and the remainder is utilized to evaluate the inference accuracy. To quantify target recommender's resistance ability, we leverage a widely-used classification metric \textit{F1 Score} to measure the performance of the attacker and show results in Figure~\ref{fig:att_f1}. Note that we set user/item dimension $d$ to 64 and use a 2-layer MLP in Eq (\ref{eq:rec}), and report inference accuracy when train-test split ratio $\zeta$ is $10\%$ and $20\%$ respectively. Correspondingly, lower F1 Score demonstrates higher resistance to this attribute inference attack. 

Firstly, we take the whole $\Delta \Theta$ (i.e., Full Version) as the input of the proposed attacker model by simply flattening all parameters from different components of a well-trained FedRec. In Figure~\ref{fig:att_f1}, we can see that the FedRec that is based on the GCN-based recommender are significantly vulnerable to our proposed attribute inference attack. Specifically, the inference attack accuracy achieves $0.528$ on age attribute and $0.677$ on gender attribute on ML-100K, while $0.522$ on age attribute and $0.684$ on gender attribute on ML-1M, though a small group of users (i.e., $\zeta = 10\%)$ are compromised. It confirms that even a fully trained deep recommender in the federated setting can leak significant amount of information about users' sensitive attributes.

Then, to understand and demonstrate the impact of model parameters derived from different components, we compare the inference accuracy of the attacker on each component of the local RedRec separately. As introduced in Section~\ref{sec:base_rec}, the local FedRec is mainly composed of three components, namely user component, item component and MLP. From Figure~\ref{fig:att_f1} (a), it is clear that these three components exhibit various degrees of information leakage. User component leaks more attribute information about each user on both two dataset, compared to the other two components. The reason behind this is twofold. By directly processing the user features, the parameters in the user embedding layer inevitably remember much more information of the original user feature, thus leaking more information. Additionally, the GCN layer aggregates the embeddings of the user's interacted items that contain other similar users' attribute information. Furthermore, though MLP does not directly contain user features, it still leaks much user information.

Due to that the recommender accuracy mainly depends on the generalization ability of the MLP, we further perform attacks with individual layers in MLP to study the vulnerability of each layer in the MLP. The results from Figure~\ref{fig:att_f1} (b) show that combining all parameters from multiple layers of MLP does not obtain significant accuracy gain. This is because of the information overlap among various layers and high dimension problem that is common in classification task. Notably, the first layer leaks more attribute information, compared to the last layer. One possible reason is that the first layer directly interacts with the concatenated embeddings of user and item. Based on the results, we propose a resistance function $l(\cdot)$ that maps each parameter to a resistance degree against attribute inference attacks.
\begin{equation}
\label{eq:idf}
    l(\theta_i) = \begin{cases}
    0, & \theta_i \in \Theta^{\mathcal{U}}\\
    1,& \theta_i \in \Theta^{\mathcal{V}}\\
    2,& \theta_i \in \Theta^{MLP_1}\\
    3,& \theta_i \in \Theta^{MLP_2}\\
    \end{cases}
\end{equation}
Therefore, the traditional assumption that the parameters from different components have the same attack vulnerability does not hold in the federated recommender, which motivates us to design an adaptive privacy-preserving federated recommender system to minimize utility loss.       
\vspace{-0.5em}
\section{Adaptive Privacy-preserving Mechanism}
In this section, we present the design of our privacy-preserving mechanism named APM that can defend against attribute inference attacks via an adaptive local differential privacy constraint where each component has a different privacy budget (i.e., noise factor $\lambda_i$). Algorithm~\ref{alg:adpRec} depicts the workflow of our adaptive privacy-preserving local training in FedRec. As the local model is trained based on a set of user-item interactions and user features, a traditional private training approach works by perturbing the model updates based on LDP techniques with fixed privacy budget for all model components before being submitted to the central server. However, assuming each component exhibits the same vulnerability against attribute inference attack would lead to suboptimal performance in both recommendation accuracy and privacy protection. To this end, we propose an adaptive privacy-preserving training mechanism. Specifically, we allocate a larger share of the privacy budget (larger noise factor) to model parameters with low attack resistances and a smaller share (smaller noise factor) to model parameters with high resistances. Thus, the privacy-preserving FedRec is able to defend against attribute inference attacks with little loss of recommendation accuracy. Despite the success of many DP- and LDP- based approaches~\cite{wei2020federated, wang2019collecting, abadi2016deep, bu2020deep} on classification task in the federated setting, most of them are not applicable in federated recommender scenarios or result in significant performance drop. Specifically, designed for modelling nonlinear relations between users and items, federated recommenders are optimized towards completely different learning objectives with much more complicated model structures where model parameters exhibit large variance. Particularly, applying strict clipping function applied to high-dimensional model parameters would lead to a large variance of the resulted model parameters and lose information contained in the original parameters. Following~\cite{qi2020privacy, wu2021fedgnn} that successfully applied Laplace Noise-based LDP approach to federated recommender system, we design the following adaptive privacy-preserving mechanism based on Laplace Noise to achieve optimal privacy protection:
\begin{equation}
\label{eq:adp}
\begin{split}
    \mathcal{M}(\Theta) &= clip(\Theta, \delta) + n,\\
    n &\sim Lap(0, \lambda_i),
\end{split}
\end{equation}
where $n$ is Laplace Noise with $0$ mean. The noise factor $\lambda_i = \frac{max_{\theta_i, \theta_{i}'}|\mathcal{M}(\theta_i)-\mathcal{M}(\theta_{i}')|}{p(\theta_i)}$ that controls the strength of Laplace Noise is determined by the adaptive privacy parameters $\epsilon_i$, and a larger $\lambda_i$ can bring better privacy protection. Specifically, given privacy parameters $\epsilon_{min}$, $\epsilon_{max}$ of each parameter and resistance function $l(\cdot)$ in Eq (\ref{eq:idf}), $p(\theta_i) = \epsilon_{min} + b \cdot l(\theta_i)$, where $b = \frac{\epsilon_{max}-\epsilon_{min}}{3}$ controls the privacy budget scoop. The function $clip(\cdot)$ is used to limit the value of each parameter with the scale of $\delta$ , and thus noise factor $\lambda_i$ is limited to $[\frac{2\delta}{\epsilon_{max}}, \frac{2\delta}{\epsilon_{min}}]$. After clip and randomization operation, it is more difficult to infer the raw user side information from the model parameters. Then each user device uploads its randomized local model parameters to the server without privacy leakage.

\begin{algorithm}[t]
\caption{Adaptive Privacy Protection Local Training for FedRec}
\KwIn{The number of local Epochs $\mathcal{T}_l$, local learning rate $\mu_1$, global model parameters $\Theta$, clipping bond $\delta$, privacy parameters $\epsilon_{min}, \epsilon_{max}$.}
\KwOut{Perturbed $\Theta_u^t$}
\textbf{Local Training Process:}\\
Initialization: $\Theta^t_u \leftarrow \Theta$ \;
\For{$t_2\leq \mathcal{T}_l$}
{
Draw a minibatch $\mathcal{B}$ \;
\For{$\mathcal{B}_i \in \mathcal{B}$}{
$\Theta^t_u \leftarrow \Theta^t_u - \mu_1 \frac{\partial \mathcal{L}^{rec}(\mathcal{B}_i, \Theta^t_u)}{\partial \Theta_u^t} $\;
}
\textbf{Add noise}:\\
$\Theta^t_u \leftarrow clip(\Theta^{t}_u, \delta)$\;
\For{$\theta_i^t \in \Theta^t_u$}{
$\lambda_i \leftarrow \frac{2\delta}{p(\theta_i^t)}$, $p(\theta_i^t) = \epsilon_{min} + \frac{\epsilon_{max}-\epsilon_{min}}{3} \cdot l(\theta_i^t)$\;
$\theta^t_i \leftarrow \theta^t_i + Lap(0,\lambda_i)$\;
}
$t_2 \leftarrow t_2 +1$\;
}

\label{alg:adpRec}
\end{algorithm}
\setlength{\textfloatsep}{1em}
\vspace{-1em}
\section{EXPERIMENTS}
In this section, we conduct experiments to verify the effectiveness of our solution on two tasks, namely privacy protection strength and recommendation effectiveness. In particular, we aim to answer the following research questions (RQs):
\begin{itemize}
    \item \textbf{RQ1:} Can our model effectively protect personal attributes in the presence of attribute inference attacks?
    \item \textbf{RQ2:} How does our method perform in the recommendation task?
    \item \textbf{RQ3:} What is the contribution of the novel adaptive privacy-protection mechanism?
    \item \textbf{RQ4:} Can our model resist attribute inference attacks that utilize different kinds of attack models?
    \item \textbf{RQ5:} what is the impact of key hyperparameters in privacy-preserving strength and recommendation effectiveness of our method?
    \item \textbf{RQ6:} How does APM perform w.r.t. different base recommenders?
\end{itemize}
\begin{table}[t!]
\caption{Features extracted from the dataset.}
  \begin{tabular}{p{8cm}}
    \hline
    \hline
\textbf{- Number of interacted products}\\
\textbf{- Number and percentage of each rating level (i.e., 1-5) given by a user}\\
\textbf{- Ratio of positive and negative ratings}: The percentage of low ratings (1 and 2) and high ratings (4 and 5) of a user.\\
\textbf{- Entropy of ratings}: It is calculated as $-\sum_{\forall r}Per_{r}\log Per_{r}$, where $Per_{r}$ is the percentage that a user gives the rating of $r$.\\
\textbf{- Median, min, max, and average of ratings}\\
\textbf{- Gender}: It is either male or female.\\
\textbf{- Occupation}: A total of 21 possible occupations are contained.\\
\textbf{- Age}: Age attribute is divided into 3 groups: under 45, over 35, and between 35 and 45.\\
\hline
\hline
\end{tabular}
\label{tab:ff}
\end{table}
\vspace{-1em}
\subsection{Datasets}
We use two publicly available datasets for evaluation, namely \textbf{ML-100K} and \textbf{ML-1M}~\cite{harper2015movielens}. ML-100k contains $10,000$ ratings from $943$ users on $1,682$ movies collected from the MovieLens website, while ML-1M is a larger dataset that contains 1 million ratings involving 6,039 users and 3,705 movies. Additionally, in each dataset, all users are associated with three private attributes, i.e., gender, age and occupation. In previous work~\cite{10.1145/3442381.3449813}, the experimental results show that the occupation attribute cannot be correctly inferred via the attribute inference attacks, compared to gender and age attributes. One possible reason is that occupation attribute is divided into 21 classes, which is hard for simple classification models to achieve acceptable accuracy. Therefore, we mainly focus on Gender and Age attributes in this work, and transfer the age and gender attributes into a $3$- and $2$-dimensional one-hot encoding vectors, following~\cite{10.1145/3442381.3449813}. Table \ref{tab:ff} provides a summary of all the user features we have used.

\vspace{-0.5em}
\subsection{Evaluation Metrics}
\textbf{Recommendation Effectiveness.} To evaluate the recommendation accuracy, we use the leave-one-out approach~\cite{10.1145/2911451.2911489} to split datasets for evaluation. Specifically, one item is preserved as ground truth for each user to construct a test set. We leverage hit ratio at rank $K$ ($Hit@K$) to measure the ratio of the ground-truth items that appear in the top-$K$ recommendation lists. Note that we use the entire negative item sets rather than the sampled subsets to compute Hit@K.

\textbf{Attribute Inference Attack Resistance.} To quantify a model's resistance ability against attribute inference attacks, we employ a widely used classification metric \textit{F1 score} to evaluate the inference accuracy of the attacker.

\vspace{-0.5em}
\subsection{Baseline Methods}
We compare our model with the following SOTA baselines. \textbf{Pure FedRec}: This is a pure GCN-based federated recommender system described in Section~\ref{sec:base_rec} without any privacy protection mechanism. \textbf{F-GERAI}: For the fair comparison, we extend the centralized recommender GERAI proposed in~\cite{10.1145/3442381.3449813} to a federated version that uses the information perturbation mechanism in DP to protect user attribute information. \textbf{F-GERAI-NL}. It is a variant of F-GERAI, which only enforces $\epsilon$-differential privacy by perturbing the objective function. \textbf{F-DPMF}: We extend DPMF (Differentially Private Matrix Factorization) proposed in~\cite{jingyu1763differentially} to a federated version F-DPMF. In its local recommender, objective perturbation is applied to make sure that the updated item embeddings satisfy differential privacy. \textbf{FedNews}~\cite{qi2020privacy}: Fednews is the first work that adopts Laplace Noise-based LDP method in federated news recommender system. In order to fit our recommendation setting where there is not textual content for items, the base recommender is replaced by our GCN-based recommender (refer to details in Section~\ref{sec:base_rec}). \textbf{FedRec-GN}: Based on Pure FedRec, we add Gaussian Noise~\cite{dwork2014algorithmic} into the uploaded parameters of each client, which can mask the original information. 
\vspace{-0.5em}
\subsection{Parameters Settings}
In FedRec, we set the latent dimension $d$, local learning rate, local batch size, local epoch to $64$, $0.001$, $32$ and $5$, respectively. Each device is assumed to contain an individual user, and 50\% and 10\% of users are randomly selected on ML-100K and ML-1M at each round. Model parameters in FedRec are randomly initialized using Gaussian distribution, which has $0$ mean and a standard deviation of $1$. In our proposed adaptive privacy-preserving mechanism, we set $\delta = 0.5$, the noise factor $\lambda \in \{0.017, 0.020, 0.025, 0.033\}$ and privacy parameter $\epsilon \in \{30, 40, 50, 60\}$.

\subsection{Privacy Protection Effectiveness (RQ1)}
\begin{table}[t!]
\centering
\caption{Performance of attribute-inference attack.}

\scalebox{0.95}{%
 \begin{tabular}{|p{1.0cm}<{\centering} | p{1.7cm}<{\centering}| p{0.95cm}<{\centering} |p{0.95cm}<{\centering} |p{0.95cm}<{\centering}| p{0.95cm}<{\centering}|} 
 \hline
 \multirow{2}{*}{Attribute}&\multirow{2}{*}{Method}&\multicolumn{2}{c|}{ML-100k}&\multicolumn{2}{c|}{ML-1M}\\
  \cline{3-6}
  &&$\zeta = 0.1$&$\zeta = 0.2$&$\zeta = 0.1$&$\zeta = 0.2$\\
 \hline
 \multirow{7}{*}{Age}&Pure FedRec&0.528&0.560&0.522&0.570\\
 &F-DPMF&0.584&0.599&0.579&0.590\\
 &F-GERAI&0.503&0.507&0.501&0.531\\
 &F-GERAI-NL&0.509&0.559&0.505&0.539\\
 &FedRec-GN&0.455&0.496&0.449&0.461\\
 &FedNews&0.446&0.480&0.439&0.479\\
 \cline{2-6}
 &ours&\textbf{0.429}&\textbf{0.470}&\textbf{0.414}&\textbf{0.434}\\
 \hline
 \hline
 \multirow{7}{*}{Gender}&Pure FedRec&0.677&0.708&0.684&0.682\\
 &F-DPMF&0.703&0.694&0.709&0.717\\
 &F-GERAI&0.639&0.648&0.627&0.640\\ 
 &F-GERAI-NL&0.657&0.658&0.634&0.665\\
 &FedRec-GN&0.575&0.642&0.627&0.634\\
 &FedNews&0.568&0.592&0.597&0.619\\
 \cline{2-6}
 &ours&\textbf{0.559}&\textbf{0.581}&\textbf{0.574}&\textbf{0.603}\\
 \hline

 \end{tabular}}
 \label{table:attack_base}
 \vspace{-0.5em}
\end{table}

Table~\ref{table:attack_base} shows the F1 Scores achieved by the attribute inference attacker described in Section \ref{sec:attacker} on all baselines. Note that lower F1 Scores show higher resistance to attribute inference attacks. Firstly, the attacker achieves higher inference accuracy on Pure FedRec than most of the recommender systems with differential privacy mechanisms, since it does not use any privacy-protection methods when uploading local model parameters to the central server. Notably, the attacker on F-DPMF achieves a better performance than FedRec, and a possible reason is that the local recommender of F-DPMF is a shallow model (i.e., Matrix Factorization) that simply represents users and items in a low dimensional latent space. Hence, massive original information can be preserved in the embeddings. Correspondingly, it is evidenced that the deep learning-based recommender can provide a stronger privacy guarantee due to the abstraction of multiple layers and complex nonlinear structure. Moreover, compared with recommenders that apply differential privacy on optimization process (i.e., F-GERAI, F-GERAI-NL and F-DPMF), the ones that make use of LDP methods (i.e., FedRec-GN, FedNews and ours) by directly adding noise into uploaded parameters show obvious superiority in defending against attribute inference attacks. The reason is that the privacy protection mechanisms utilized in those optimization perturbation methods cannot yield the same strength as the LDP-based methods in preventing the disclosure of sensitive information from model parameters. This also confirms that the adoption of DP-based approaches may preclude directly leaking private attributes, but these methods are unable to effectively perform higher privacy protection in the federated setting. Notably, compared with other LDP-based recommender systems, it is clear that our method yields the best performance in obscuring users' private attribute information. Finally, we can see that F-GERAI outperforms F-GERAI-NL in terms of F1 Score, due to the dual-stage perturbation setting where a relatively strong privacy protection method is applied for user feature perturbation. Meanwhile, our method can constantly achieve better results without an extra privacy budget on original features, indicating our method endows the uploaded local model parameters a stronger privacy guarantee.   
\vspace{-0.5em}
\subsection{Recommendation Effectiveness (RQ2)}
\begin{figure}[t!]
\centering
 \includegraphics[scale=0.42]{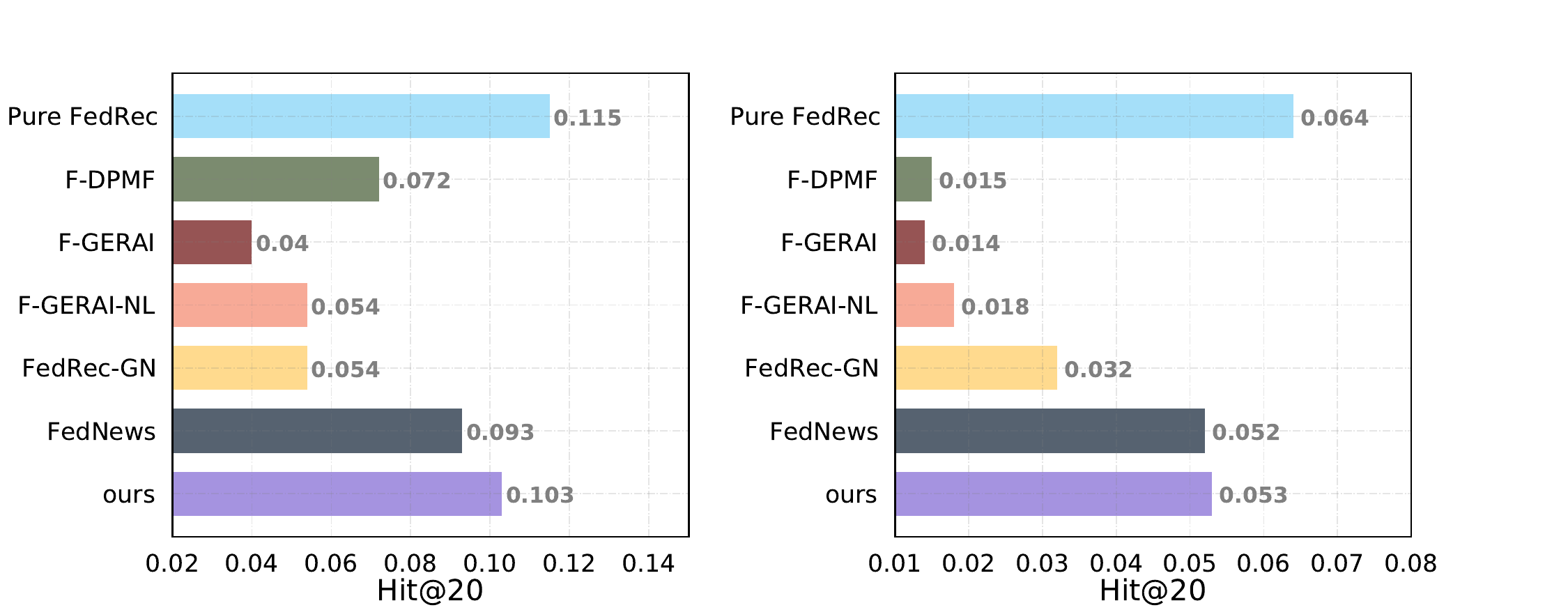}
\vspace{-0.5em}
\caption{Recommendation effectiveness results.}
\label{fig:att_acc}
\vspace{-1em}
\end{figure}

\begin{figure*}[t!]
\centering
\begin{tabular}{cccc}
\multicolumn{4}{c}{ \includegraphics[scale=0.45]{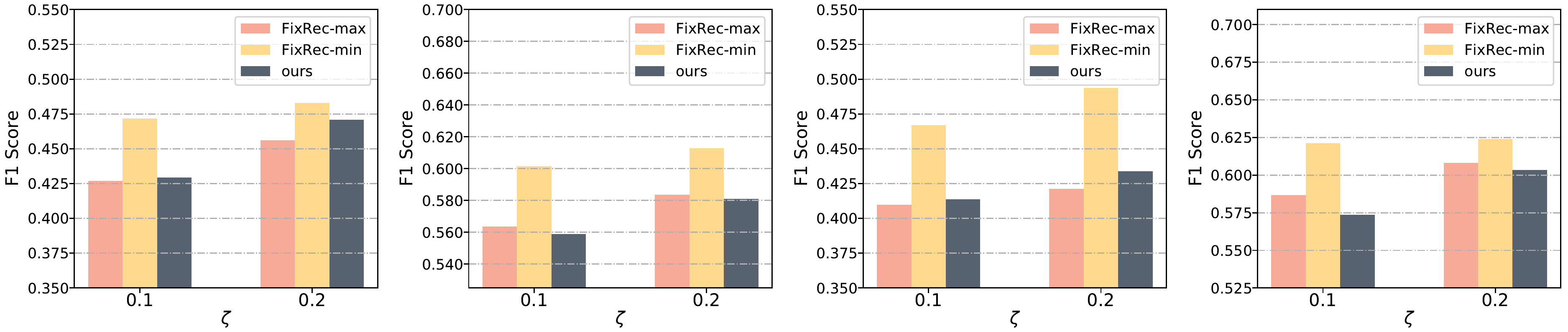}}\\
\hspace{2.8em}(a)Age on ML-100K&\hspace{2.2em}(b)Gender on ML-100K&\hspace{2.5em}(c)Age on ML-1M&\hspace{2.0em}(d)Gender on ML-1M\\
\end{tabular}
\caption{Evaluation performance of adaptive privacy-preserving mechanism against attribute inference attacks with two privacy-preserving recommenders with fixed privacy budget.}
\label{fig:att_dp}
\vspace{-1em}
\end{figure*}

         
         
  
Recommendation accuracy is an important metric in the evaluation of privacy-preserving recommenders, since protecting user privacy is usually at the expense of their recommendation accuracy. Hence, a practical recommender should resist inference attacks without sacrificing high-quality recommendations. We report all methods' performance on personalized recommendation w.r.t. Hit@20 in Figure~\ref{fig:att_acc}, and higher Hit@20 values imply higher recommendation quality. Clearly, recommendation methods that make use of privacy protection mechanisms have significant performance disparity in terms of the Hit@20. Our method outperforms all privacy-preserving baselines by a large margin in both datasets, thanks to our proposed adaptive perturbation mechanism in which the values of privacy budget are dynamically adjusted according to the vulnerability of each model component. Furthermore, LDP-based recommenders achieve significantly better results than DP-based ones while maintaining stronger privacy protection, which further confirms the effectiveness of LDP-based mechanisms in the federated scenarios. Although F-GERAI can achieve promising recommendation performance in the traditional centralized setting~\cite{10.1145/3442381.3449813}, it does not fit well the federated setting, as the way it adds noise to the side information is specialized for the centralized recommendation, and it is harmful to the federated recommendation. Compared with FedRec-GN, Fednews achieves higher recommendation accuracy, which implies that Laplace Noise can ensure recommendation effectiveness while avoiding breaching users' privacy. This is the reason that we adopt it as the basic privacy protection mechanism in our method.
\vspace{-0.5em}
\subsection{Importance of Adaptive Privacy Mechanism (RQ3)}

To better understand the benefits brought by our proposed adaptive privacy mechanism, we compare with two variants of our method, FixRec-max and FixRec-min that hold fixed privacy budgets on ML-100K and ML-1M. Setting an appropriate value for the privacy budget $\lambda$ is crucial, as it controls the trade-off between privacy protection level and recommendation accuracy. To explore the suitable range of $\lambda$ in Eq (\ref{eq:adp}), we first fix privacy parameter $\epsilon$ for all model components and then vary its value while keeping the other hyperparamerters unchanged. From the experimental results, we find that the value of $\epsilon$ should be limited to $E = [40, 60]$ and thus $\lambda$ should be within the limits of $\mathcal{H} = [0.017, 0.033]$. A larger $\lambda$ can cause a severe performance drop on recommendation accuracy, while the recommender with a too small value of $\lambda$ fails to obscure users' private attribute information. Hence, we set the privacy budget $\lambda$ separately for each component according to their resistances obtained in Section~\ref{sec:attacker}, that is $0.017$ ($\epsilon = 60$) for Item Component, $0.025$ ($\epsilon = 40$) for first layer of MLP, $0.020$ for second layer of MLP ($\epsilon = 50$) and $0.033$ ($\epsilon = 30$) for User Component. Then we report the attribute inference results and recommendation results of our method and the two variants that respectively adopt the minimum fixed privacy budget (named as FixRec-min) and the maximum fixed privacy budget (named as FixRec-max) of $\mathcal{H}$. 

From Figure~\ref{fig:att_dp}, we can see that FixRec-max achieves the best performance in defending against attribute inference attacks, since a larger $\lambda$ requires a larger amount of noise to be injected into the model parameters, leading to the larger information obfuscation. It is worth mentioning that our method not only outperforms FixFed-min but also achieves comparable results with Fixed-max, indicating our adaptive privacy mechanism can still effectively provide satisfactory privacy level with a lower privacy budget. 

Table~\ref{fig:dp_acc} shows that FixRec-min outperforms FixRec-max by a large margin, especially on ML-100K dataset. The reason is that a larger amount of noise is injected to the training process of FixRec-max, which negatively influences the recommendation accuracy. Furthermore, our model significantly outperforms FixRec-max and yields recommendation results that are close to the FixRec-min, indicating that an adaptive privacy budget can be beneficial to significantly reduce the utility loss that is inevitably caused by the LDP-based approaches. Notably, our method is able to achieve comparable recommendation results as FixRec-min and resistance ability as FixRec-max. It confirms the effectiveness of our proposed adaptive privacy mechanism, which helps our federated recommender system resist attribute inference attack and avoid unnecessary utility loss.
\subsection{Robustness against Different Attribute Inference Attackers (RQ4)}
\begin{table}[t]
    \caption{Recommendation results of our model with adaptive privacy budget and two variants with fixed privacy budget.}
    \centering
    \scalebox{1}{%
    \begin{tabular}{|p{3cm}<{\centering} |p{2cm}<{\centering}|p{2cm}<{\centering}|}
        \hline
         Method&ML-100K&ML-1M\\
          \cline{1-3}
          FixRec-max&0.067&0.043\\
          FixRec-min&0.105&0.054\\
          \hline
          ours&0.103&0.053\\
         \hline
         
         
    \end{tabular}}
    \vspace{-0.5em}
    \label{fig:dp_acc}
  
\end{table}
\begin{table*}[t]
\centering
\caption{Performance of attribute-inference attack w.r.t. different types of attacker.}

\scalebox{1}{%
 \begin{tabular}{|p{1.6cm}<{\centering} | p{2.2cm}<{\centering}| p{1.2cm}<{\centering} |p{1.2cm}<{\centering} |p{1.2cm}<{\centering}| p{1.2cm}<{\centering}|p{1.2cm}<{\centering} |p{1.2cm}<{\centering} |p{1.2cm}<{\centering}| p{1.2cm}<{\centering}|p{1.2cm}<{\centering}|p{1.2cm}<{\centering}|} 
 \hline
 \multirow{2}{*}{Attribute}&\multirow{2}{*}{Method}&\multicolumn{4}{c|}{ML-100k}&\multicolumn{4}{c|}{ML-1M}\\
  \cline{3-10}
  &&DT&SVC&KNN&AIA&DT&SVC&KNN&AIA\\
 \hline
 \multirow{7}{*}{Age}&Pure FedRec&0.427&0.465&0.491&0.528&0.460&0.487&0.520&0.522\\
 &F-DPMF&0.417&0.401&0.512&0.584&0.443&0.390&0.449&0.579\\
 &F-GERAI&0.406&0.422&0.483&0.503&0.415&0.408&0.425&0.501\\
 &F-GERAI-NL&0.413&0.474&0.488&0.509&0.419&0.449&0.430&0.505\\
 &FedRec-GN&0.399&0.380&0.446&0.455&0.397&0.381&0.445&0.449\\
 &FedNews&0.401&0.406&0.417&0.446&0.388&0.390&\textbf{0.417}&0.439\\
 \cline{2-10}
 &ours&\textbf{0.394}&\textbf{0.356}&\textbf{0.413}&\textbf{0.429}&\textbf{0.377}&\textbf{0.333}&0.428&\textbf{0.414}\\
 \hline
 \hline
 \multirow{7}{*}{Gender}&Pure FedRec&0.601&0.566&0.677&0.677&0.671&0.640&0.680&0.684\\
 &DPMF&0.580&0.559&0.656&0.703&0.572&0.520&0.647&0.709\\
 &F-GERAI&0.571&0.566&0.613&0.639&0.612&0.559&0.599&0.627\\
 &F-GERAI-NL&0.580&0.587&0.646&0.657&0.618&0.566&0.631&0.634\\
 &FedRec-GN&0.564&0.481&0.625&0.575&0.616&0.520&0.601&0.627\\
 &FedNews&0.561&0.474&\textbf{0.608}&0.568&0.614&0.513&0.596&0.597\\
 \cline{2-10}
 &ours&\textbf{0.556}&\textbf{0.472}&0.613&\textbf{0.559}&\textbf{0.550}&\textbf{0.496}&\textbf{0.550}&\textbf{0.574}\\
 \hline

 \end{tabular}}
 \label{table:attack_type}
 \vspace{-0.5em}
\end{table*}
In real-life scenarios, they are many available models that can be selected to perform attribute inference attacks for the adversary, so the attack models are usually unknown and unpredictable. To better understand the vulnerability of our method and other comparable methods, we design several different types of attack models, namely Decision Tree (DT), SVC and KNN, that are frequently adopted approaches in classification tasks. In this study, we use the full version of $\Delta \Theta$ that is derived from well-trained federated recommenders for all attackers and set $\zeta = 0.1$. Table~\ref{table:attack_type} shows the attribute inference accuracy of each attacker. It is obvious that our proposed method outperforms all the comparison methods in most scenarios, which implies that our method can more effectively defend against attribute inference attacks and provide a stronger privacy guarantee when confronted with unknown attacker models. Though Fednews achieves slightly better results when attacker is a KNN-based model, it falls behind our model in all other cases and yields inferior performance in recommendation task. Furthermore, the FedRec-GN cannot perform as well as Fednews that uses Laplace Noise, which further verifies the advantages of utilizing Laplace Noise-based LDP method in defending against inference attacks. Finally, DNN-based attacker (i.e., AIA) outperforms other attackers in most scenarios, since its superiority in learning non-linear correlation between input features and target labels.

\vspace{-0.5em}
\subsection{Parameter Sensitivity (RQ5)}
\begin{figure}[t]
\centering
\begin{tabular}{cc}
\multicolumn{2}{c}{ \includegraphics[scale=0.40]{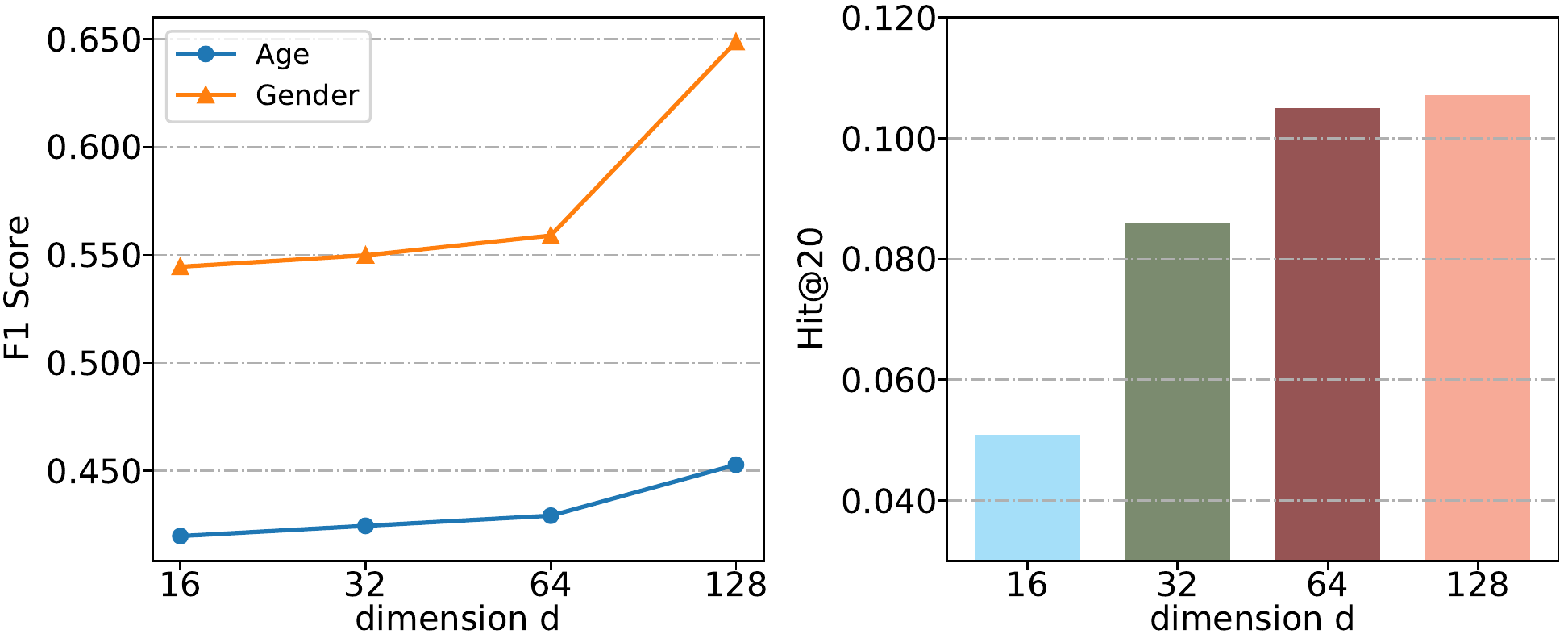}}\\
\end{tabular}
\vspace{-0.5em}
\caption{Inference attack and recommendation results on ML-100K w.r.t. dimension $d$}
\vspace{-0.5em}
\label{fig:d_sens}
\end{figure}

We answer RQ5 by investigating the performance fluctuations of our method with varied hyperparameters, particularly embedding dimension $d$ and train-test split ratio $\zeta$. Due to the space limitation, we only showcase the results on ML-100K dataset, and similar results are also achieved on ML-1M dataset. Specifically, we tune the value of $d$ or $\zeta$ while keeping the other hyperparameters unchanged, and record the new recommendation accuracy and inference attack results achieved in Figure~\ref{fig:d_sens} and Table~\ref{fig:d_sens1}. 

\textbf{Dimension $d$}. 
We examine the value of dimension $d$ in $\{16, 32, 64, 128\}$. In general, the dimension $d$ mainly controls the models' expressiveness. Obviously, the recommendation accuracy of our model benefits from a relatively larger dimension $d$, and then the performance gain appears to become less significant when $d$ reaches $64$. In the attribute inference task, our method with a smaller $d$ shows higher resistance to the inference attack. One possible reason is that the core of a classification model is the ability to assign a class to an object based on input features. A larger $d$ can increase the model's expressiveness and thus the attacker can learn more information from the model parameters. Fortunately, our model is able to achieve competitive performance in both two tasks when $d = 64$. 

\textbf{Train-test split ratio $\zeta$}. Table~\ref{fig:d_sens1} shows attack accuracy for different types of attackers with varied prior knowledge (i.e., the size of training data). As expected, increasing the size of the attackers' training dataset improves the accuracy of the attribute inference attacks. It is worth mentioning that though the attackers collect much more prior knowledge (i.e., $\zeta = 30\%$), our model still achieves acceptable performance in resistance to attribute inference attacks. 

\subsection{Applications of APM (RQ6)}
As one of the main contributions of this paper is that we propose APM, an adaptive privacy-preserving mechanism to defend against attribute inference attacks without sacrificing high-quality recommendation results, we conduct a comparison between APM and other privacy-preserving mechanisms in two most representative recommenders (i.e., LightGCN~\cite{he2020lightgcn} and NCF~\cite{he2017neural}) under the federated learning protocol (described in Section \ref{sec:fl}). Specifically, LightGCN is widely used in centralized setting since it offers simplicity via the omission of excessive nonlinear components, while NCF is generic and can generalize various centralized recommenders under its framework. Table \ref{tab:application} reports experimental results in recommendation and attribute inference tasks. Note that model without suffix, with +LDP and +APM mean that it is designed without any privacy protection mechanism, with normal Laplace Noise-based LDP and our proposed APM approach respectively. From the results, we can see that base recomenders that make full use of our APM consistently outperform other mechanisms in both two tasks. The results confirm that our proposed APM can seamlessly integrate with most main-stream federated recommenders to protect user privacy better while costing less recommendation accuracy. 
\begin{table}[t]
    \caption{Inference attack results on ML-100K w.r.t. train-test split ratio $\zeta$.}
    \centering
    \scalebox{1}{%
    \begin{tabular}{|p{1.2cm}<{\centering} |p{0.8cm}<{\centering}|p{1.0cm}<{\centering}|p{1.0cm}<{\centering}|p{1.0cm}<{\centering}|p{1.0cm}<{\centering}|}
        \hline
        \multirow{2}{*}{Attribute}&\multirow{2}{*}{$\zeta$}&\multicolumn{4}{c|}{Attack Models (F1 Score)}\\
        \cline{3-6}
        &&AIA&DT&SVC&KNN\\
         \hline
         \multirow{4}{*}{Age}&10\%&0.429&0.394&0.356&0.413\\
         &20\%&0.470&0.403&0.361&0.430\\
         &30\%&\textbf{0.498}&\textbf{0.403}&\textbf{0.382}&\textbf{0.436}\\
         \hline
         \hline
         \multirow{4}{*}{Gender}&10\%&0.559&0.556&0.472&0.613\\
         &20\%&0.581&0.570&\textbf{0.483}&0.623\\
         &30\%&\textbf{0.585}&\textbf{0.572}&0.482&\textbf{0.627}\\
         \hline
    \end{tabular}}
    \label{fig:d_sens1}
  
\end{table}

\begin{table}[t!]
\centering
\caption{Performance of privacy-preserving mechanisms in different federated recommenders.}
\scalebox{0.95}{%
 \begin{tabular}{|p{1.2cm}<{\centering} | p{2.5cm}<{\centering}| p{1.0cm}<{\centering} |p{1.0cm}<{\centering} |p{1.0cm}<{\centering}|} 
 \hline
 \multirow{2}{*}{Dataset}&\multirow{2}{*}{Method}&\multirow{2}{*}{Hit@20}&\multicolumn{2}{c|}{F1 Score}\\
  \cline{4-5}
  &&&Age&Gender\\
 \hline
 \multirow{6}{*}{ML-100K}&FedLightGCN&0.080&0.594&0.696\\
&FedLightGCN+LDP&0.073&0.462&0.585\\
&FedLightGCN+APM&0.076&0.415&0.573\\
\cline{2-5}
&FedNCF&0.075&0.509&0.651\\
&FedNCF+LDP&0.070&0.488&0.637\\
&FedNCF+APM&0.073&0.474&0.620\\
\hline
\hline
\multirow{6}{*}{ML-1M}&FedLightGCN&0.082&0.550&0.728\\
&FedLightGCN+LDP&0.077&0.471&0.601\\
&FedLightGCN+APM&0.080&0.450&0.585\\
\cline{2-5}
&FedNCF&0.076&0.509&0.676\\
&FedNCF+LDP&0.072&0.460&0.675\\
&FedNCF+APM&0.074&0.422&0.658\\
\hline
 \end{tabular}}
 \label{tab:application}
\end{table}

\section{RELATED WORK}
  

\textbf{Attribute Inference Attacks and defenses.} 
Attribute inference attacks aim to infer users' sensitive information by carefully designing an attacker model based on the collected information such as outputs and structure of the target model, and training dataset.~\cite{lindamood2009inferring,he2006inferring,gong2014joint} infer attributes information by incorporating available target users' friend information. Behavior-based approaches construct attack models based on users' behavioral data (e.g., movie-rating behavior~\cite{weinsberg2012blurme} and Facebook likes~\cite{kosinski2013private}).~\cite{jia2017attriinfer,gong2018attribute,gong2016you} achieve adversarial purpose by leveraging both users' friend and behavior information. In the centralized recommendation context, attribute inference attacks attempt to infer users' private information (e.g., demographic features) from publicly available information (i.e. recommendation results and user-item interaction data). To address such privacy issues, there have been emerging research efforts on developing privacy-preserving centralized recommender systems~\cite{10.1145/3442381.3449813, beigi2020privacy, wang2020next}. RAP is proposed to enhance attack-resistance of conventional recommender systems by utilizing an adversarial learning paradigm where a recommender model and a pre-defined attack model are trained against each other. But the design of RAP makes it effectively defend against a specific attacker model. Another variant is encryption-based methods that use encryption techniques such as homomorphic encryption~\cite{kim2018efficient,chai2020secure}. However, in these approaches, an extra third-party crypto-service provider is required, so they are computation intensive. Recently, differential privacy becomes a well-established technique to address privacy issues, since it can provide a mathematically provable guarantee~\cite{mcsherry2009differentially,berlioz2015applying,liu2015fast}. For example, GERAI~\cite{10.1145/3442381.3449813} is proposed to perturb the user's side information and optimization process to prevent privacy leakage via recommendations. However, the centralized recommender systems that store and train users' personal data centrally are still suffering from enormous and unprecedented privacy issues.

\textbf{Federated Learning.} To tackle privacy issues existing in centralized scenarios, a common practice is to deploy the online system in a federated setting, which enables users to collaboratively learn a global model while keeping all the sensitive data on local devices~\cite{yuan2023interaction, yuan2023federated}. Federated learning starts training by initializing a shared global model, then a subset of existing clients is selected to train their local models based on the private dataset and submit the updated model parameters. With these updates, the server operates aggregation of the received updates to replace the parameters of the global model. There are some works that attempt to develop federated recommender systems.~\cite{ammad2019federated} aims to bind Matrix Factorization approach into the federated setting. In FedFast~\cite{muhammad2020fedfast}, a novel sampling method that selects participating clients in each training iteration and an activate aggregation rule that combines locally trained models are devised to reduce the communication cost and speed up the convergence rate of current federate recommender systems.~\cite{chen2018federated} applies meta learning in the federated model with a shared meta learner, which differs from the conventional FL setting which shares a global model. Though federated recommender can achieve satisfactory recommendation accuracy without accessing users' data, recent works show that it is yet to provide a privacy guarantee of users' privacy. In~\cite{wang2021fast}, the central server adopts a DP-based mechanism to perturb global recommender, which can defend against attacks from malicious participants who can infer private information via the shared global model. However, it is only effective when the central server is in a sterile environment. In real-life scenarios, the server is curious about inferring the users' private attributes via received updates, leaving a backdoor for attribute inference attacks. Hence, privacy protection methods should be applied to each individual's local model parameters before sending to the server. To enhance privacy protection,~\cite{qi2020privacy,wu2021fedgnn} applied LDP-based approaches on the local model parameters. However, the existing privacy-preserving works in federated learning assume all components exhibit the same resistance degree, which is violated and causes a severe recommendation performance drop. These limitations motivated us to propose an adaptive privacy-preserving recommender system that is able to counter attribute inference attacks in the federated setting, while maintaining high recommendation accuracy.
\section{Conclusion And Future Work}
In this paper, we focused on the privacy issues of federated recommenders confronted with attribute inference attacks. To provide a comprehensive analysis of current federated recommender systems, we design a novel attribute inference attacker to show the vulnerability of each internal component of the recommender. In accordance with the experimental results, we proposed an adaptive privacy-preserving federated recommender system to protect users' sensitive data in defending against inference attacks while maintaining high-quality recommendation results. Specifically, to minimize utility loss caused by LDP-based approaches, we improve the naive LDP mechanisms with an adaptive privacy budget based on the resistance degree. The experimental results validate the superiority of our solution by comparing with the baseline approaches. In the future work of privacy-preserving federated recommenders, it will be appealing to further investigate privacy protection against active attackers that can participate in the training of federated recommender systems and craft adversarial parameter updates for follow-up attribute inference attacks.


%

\ifCLASSOPTIONcompsoc
  \section*{Acknowledgments}
\else
  \section*{Acknowledgment}
\fi

This work is supported by Australian Research Council Future Fellowship (Grant No. FT210100624), Discovery Project (Grant No. DP190101985).

\ifCLASSOPTIONcaptionsoff
  \newpage
\fi

\bibliographystyle{IEEEtran}
\bibliography{bibreference.bib}
\vspace{-4em}
\begin{IEEEbiography}[{\includegraphics[width=1in,height=1.25in,clip,keepaspectratio]{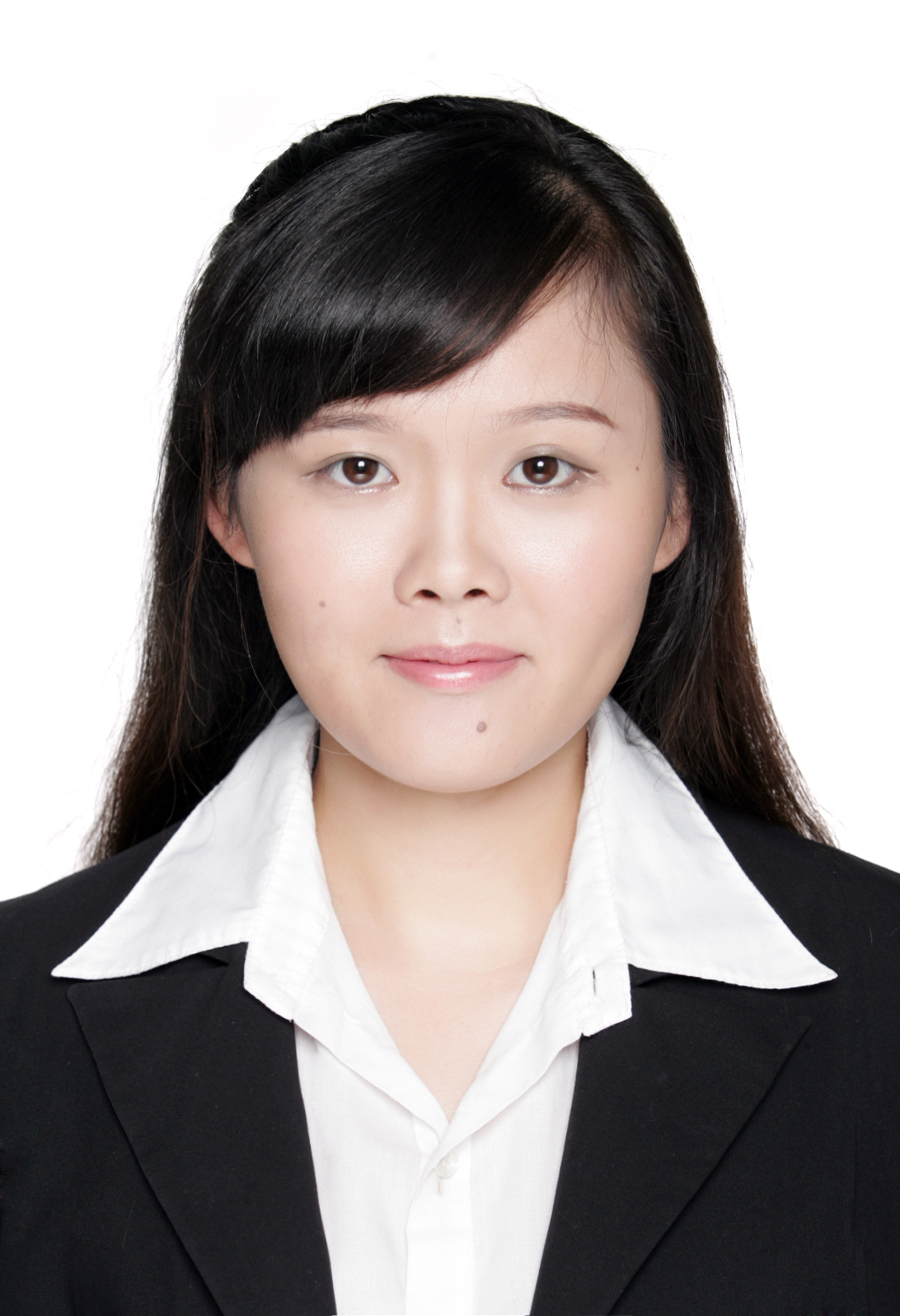}}]{Shijie Zhang} is currently a computer science Ph.D. student at the School of Information Technology and Electrical Engineering, The University of Queensland. She obtained Master Degree from The University of Queensland in 2018 and Bachelor Degree from Shandong University. Her research interests include data mining, recommender system, deep learning, and federated learning.
\end{IEEEbiography}
\vspace{-25em}
\begin{IEEEbiography}[{\includegraphics[width=1in,height=1.25in,clip,keepaspectratio]{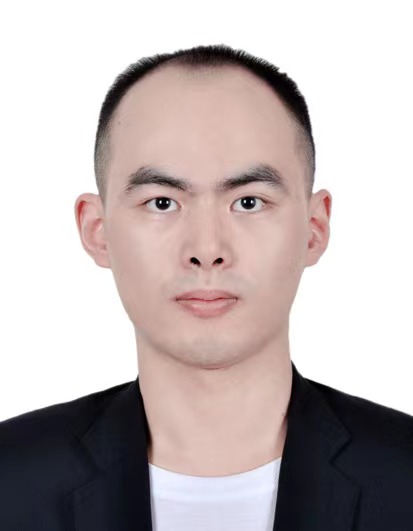}}]{Wei Yuan} is a computer science PhD. student at the School of Information Technology and Electrical Engineering, The University of Queensland. He obtained Master Degree from Nanjing University. His research focuses on trustworthy and secure recommender systems, automatic program repair, and natural language generation.
\end{IEEEbiography}
\vspace{-25em}
\begin{IEEEbiography}[{\includegraphics[width=1in,height=1.25in,clip,keepaspectratio]{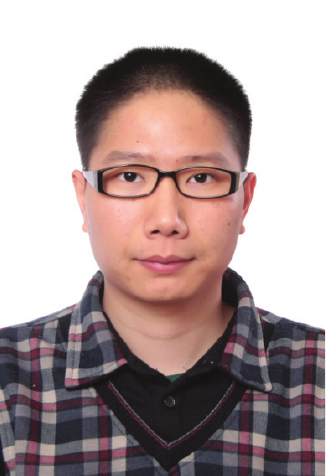}}]{Hongzhi Yin} works as an ARC Future Fellow, associate professor, and director of the Responsible Big Data Intelligence Lab (RBDI) at The University of Queensland, Australia. He has made notable contributions to predictive analytics, recommendation systems,  graph learning, social media analytics, and decentralized and edge intelligence. He has received numerous awards and recognition for his research achievements. He has been named to IEEE Computer Society’s AI’s 10 to Watch 2022 and Field Leader of Data Mining \& Analysis in The Australian's Research 2020 magazine.  In addition, he has received the prestigious Australian Research Council Future Fellowship 2021, the Discovery Early Career Researcher Award 2016,  Research.com Rising Star of Science Award 2022, 2023 and 2022 AI 2000 Most Influential Scholar Honorable Mention in Data Mining.  He was featured among the 2022 and 2021 World's Top 2\% Scientists Lists (Career Impact) published by Stanford University. He has published 240+ papers with H-index of 58, including 140+ CCF A and 70+ CCF B, 140+ CORE A* and 70+ CORE A.  His research has won 8 international and national Best Paper Awards, including Best Paper Award - Honorable Mention at WSDM 2023, Best Paper Award at ICDE 2019, Best Student Paper Award at DASFAA 2020, Best Paper Award Nomination at ICDM 2018,  ACM Computing Reviews' 21 Annual Best of Computing Notable Books and Articles, Best Paper Award at ADC 2018 and 2016,  and Peking University Distinguished Ph.D. Dissertation Award 2014. 
\end{IEEEbiography}
\end{document}